\documentclass[aps,prd,article]{revtex4}
\usepackage{amsmath}

\newcommand{\avg}[1]{\left< #1 \right>} 
\newcommand{\ket}[1]{\left| #1 \right>} 
\newcommand{\bra}[1]{\left< #1 \right|} 
\newcommand{\braket}[2]{\left< #1 \vphantom{#2} \right|
 \left. #2 \vphantom{#1} \right>} 

\begin{document}

\title{Chirality-energy conversion induced by static magnetic effects on free electrons in quantum field theory}
\author{P. Kurian}
\email{pkurian@howard.edu}
\affiliation{Quantum Biology Laboratory, Howard University, and Department of Medicine, Howard University College of Medicine, Washington DC, USA}

\begin{abstract}
Magnetic effects on free electron systems have been studied extensively in the context of spin-to-orbital angular momentum conversion. Using a quantum field theory framework, we derive a similar relationship in the non-relativistic limit for the energy of electrons with momentum directed along the axis of a spatiotemporally constant, weak magnetic field. For a single electron the expectation value of the maximum energy shift, which is fixed by our defined \textit{chirality index} of the electron state, is computed perturbatively to first order as ${\sim}15\%$ of the electron rest mass. This effect is orders of magnitude larger than that predicted by the quantum mechanical Zeeman shift. We then show, in the low-mass approximation, an analogous conversion between energy and chirality for a system of free electrons and suggest possible experimental tests of this phenomenon in electron states encountered across multiple physics disciplines.
\end{abstract}

\maketitle
The effects of a magnetic field on the free-electron spin have been computed recently \cite{PLA2016} in the framework of quantum field theory (QFT), suggesting potential applications for spin-to-orbital angular momentum conversion in spintronic devices and electron vortex beams \cite{Ultramicro}. The work described here is motivated by the rich history in condensed matter and particle physics of extending quantum mechanical results by application of QFT. In particular, our study makes use of the fact that in QFT spin is precisely defined at the outset as a function of the quantum fields, rather than arising in quantum mechanics as an ad-hoc addition to the orbital angular momentum. Starting from the expression of the spin $\vec{S} = (S_1, S_2, S_3) = (S_{23}, S_{31}, S_{12})$ in terms of the fields,
\begin{equation} \label{Spin}
S_{ab} = \int d \mathbf{x} \, \psi^\dagger (\mathbf{x}) \, \frac{i}{2} \gamma^a \gamma^b \, \psi(\mathbf{x}),
\end{equation}
the complete expression for the spin shift caused by the magnetic field \cite{PLA2016} is given by 
\begin{equation}
\Delta_{\vec{A}} \vec{S} = |e| \int d\vec{x} \, \, [\vec{A} \times \vec{\rho}_E],
\end{equation}
where $\vec{A}$ is the magnetic potential and $\vec{\rho}_E = \frac{i}{m_e}\psi^\dagger \vec{\gamma} \psi$.

The purpose of the present letter is to compute the analogous effect on the energy of a free electron state. With this aim, we shall begin from the explicitly hermitian expression for the Hamiltonian of a free electron, given in terms of the four components $\psi_1, \psi_2, \psi_3, \psi_4$ of the electron field $\psi(\mathbf{x})$:
\begin{eqnarray}
\mathcal{H} = \int d\vec{x} && \left[ -\frac{i}{2}\overline{\psi} \gamma^{p} \partial_{p} + \frac{i}{2} (\partial_{p}\overline{\psi})\gamma^{p} + m_e \overline{\psi} \right] \psi \nonumber\\
= \int d\vec{x} && \left[2m_e \, \Re\left(\, \psi_1^\ast \psi_3 + \psi_2^\ast \psi_4\,\right) \right. - \left.  \,\Im\left(\, \psi_1^\ast \partial_3 \psi_1 +\psi_1^\ast \left(\, \partial_1 -i\partial_2\,\right)\psi_2 \right.\right. +\left.\left. \psi_2^\ast \left(\, \partial_1 +i\partial_2\,\right)\psi_1 -\psi_2^\ast \partial_3 \psi_2  \,\right)\right.+  \nonumber\\
&&\left. \,\Im\left(\, \psi_3^\ast \partial_3 \psi_3 +\psi_3^\ast \left(\, \partial_1 -i\partial_2\,\right)\psi_4\right. \right. + \left.\left. \psi_4^\ast \left(\, \partial_1 +i\partial_2\,\right)\psi_3 -\psi_4^\ast \partial_3 \psi_4  \,\right)
\right.].
\label{ham}
\end{eqnarray}
It is clear above that the implied summation over $p$ indexes the three spatial components only. The introduction of an electromagnetic potential $A_{\mu}=(A_0, \vec{A})$ will modify the free Hamiltonian following the conventional Dirac prescription $\partial_{\mu} \rightarrow \partial_{\mu} - i |e| A_{\mu}$. Keeping only those terms to first order in the components of $A_{\mu}$, we find the energy shift due to the electromagnetic potentials:
\begin{equation}
\Delta_{A_{\mu}} \mathcal{H} = 2 |e| \int d\vec{x} \, \left\{A_0 \rho_0 + \vec{A} \cdot \left[\Re \left(-\psi_1^*\psi_2+\psi_3^*\psi_4\right), \Im \left(-\psi_1^*\psi_2+\psi_3^*\psi_4 \right), \frac{1}{2} \left(-|\psi_1|^2 + |\psi_2|^2 + |\psi_3|^2 - |\psi_4|^2 \right) \right] \right\},
\end{equation}
where $\rho_0 = \sum_{j=1}^4 |\psi_j|^2$.

In this letter, we will only be interested in the magnetic shift produced by the vector potential $\vec{A}$. While our Equation (\ref{ham}) is derived from a Lagrangian that is manifestly gauge invariant from the beginning, our ensuing results will not all be written in a covariant form. Indeed, because we have restricted our scope to the contributions from magnetic fields only, several equations below will not be manifestly covariant. The shift derived from the electric fields ($\vec{E} = -\nabla A_0$) is thus not included. However, in the non-relativistic limit (NRL), where $\psi_{1}, \psi_{2} \gg \psi_{3}, \psi_{4}$, the expression for the magnetic shift becomes particularly simple:
\begin{equation}
\Delta_{\vec{A}} \mathcal{H}_{NRL} = -2 |e| \int d\vec{x} \, \vec{A} \cdot \vec{s}_{NRL},
\label{nrl}
\end{equation}
where $\vec{s}_{NRL}$ is the spin current in the NRL. The spin current is defined generally as the quantity whose integral gives the spin vector itself, $\vec{S}=\int d\vec{x} \, \vec{s}$, and its components are given by
\begin{eqnarray}
\nonumber s_1=&& \Re\left(\,\psi_1^\ast \psi_2 + \psi_3^\ast \psi_4\, \right)\label{s1}\\ 
\nonumber s_2=&& \Im\left(\,\psi_1^\ast \psi_2 + \psi_3^\ast \psi_4\, \right)\label{s2}\\ 
s_3=&& \frac{1}{2}\left(\,|\psi_1|^2 - |\psi_2|^2 \right. + \left. |\psi_3|^2 - |\psi_4|^2 \,\right).
\label{s3}
\end{eqnarray}
A relevant feature of the obtained result is that such a magnetic energy shift is different from what one would expect from a quantum mechanical description. As is well known, this expression---called the Zeeman effect---can be written for an electron as a scalar product between the electron angular momentum $\vec{J}$ and the magnetic field $\vec{B}$. For the contribution coming from the electron spin $\vec{S}$, we would have in quantum mechanics
\begin{equation}
\Delta _{\vec{A}} \mathcal{H}_{Zeeman} = -g_s \mu_B \vec{B} \cdot \vec{S},
\label{Zeeman}
\end{equation}
where $g_s \approx 2$ is the gyromagnetic factor and $\mu_B=|e|/2m_e$ is the Bohr magneton. One can see a faint resemblance between Equation (\ref{Zeeman}) above and the QFT expression of Equation (\ref{nrl}), with roughly a replacement of the magnetic field $\vec{B}$ with the magnetic potential $\vec{A}$ and likewise of the spin $\vec{S}$ with the spin current $\vec{s}_{NRL}$.  

We now want to compute the expectation value of the energy shift in Equation (\ref{nrl}) in a specific one electron state. Consider for simplicity the state with momentum $\vec{k}$ along the $z$ axis and defined as a linear combination of the two spin eigenstates with complex coefficients:
\begin{equation} \label{Psi}
\left |\Psi(\vec{k}) \right \rangle = \lambda_+ \left|\uparrow, \vec{k} \right \rangle + \lambda_- \left|\downarrow,\vec{k}\right \rangle, 
\end{equation}
where the spin eigenstates are given by
\begin{equation}
\left|\uparrow, \vec{k} \right \rangle = \sqrt{2E} \, a_{\vec{k}}^{\uparrow\,\dagger} |0\rangle, \quad \left|\downarrow, \vec{k} \right \rangle = \sqrt{2E} \, a_{\vec{k}}^{\downarrow\,\dagger} |0\rangle, 
\label{spinup/down}
\end{equation}
with $E=\sqrt{m_e^2 + |\vec{k}|^2}$ and the state normalization fixed according to the prescription of Peskin and Schroeder \cite{Peskin} as
\begin{equation}
\braket{\uparrow, \vec{k}\,\,\,}{\uparrow, \vec{k}} = \braket{\downarrow, \vec{k}\,\,\,}{\downarrow, \vec{k}} = 2E (2\pi)^3 \delta^{(3)}(0).
\label{spinnorm}
\end{equation}
The normalization of our single electron state follows immediately:
\begin{equation}
\braket{\Psi(\vec{k})\,\,}{\Psi(\vec{k})} =  2E (2\pi)^3 \delta^{(3)}(0) \left(|\lambda_+|^2 + |\lambda_-|^2 \right).
\end{equation}

The starting expression for the expectation value of the energy shift is
\begin{equation}
\avg{\Delta_{\vec{A}} \mathcal{H}_{NRL}}  = \frac{\bra{\Psi(\vec{k})}\Delta_{\vec{A}} \mathcal{H}_{NRL}\ket{\Psi(\vec{k})}}{\braket{\Psi(\vec{k})}{\Psi(\vec{k})}}. \label{expavgH}
\end{equation}
To compute it, one needs to introduce the Fourier transforms of the electron fields, limiting the expressions to only those containing the fermionic creation and annihilation operators:
\begin{equation}
\begin{gathered}
\psi(\mathbf{x}) = \int \frac{d\vec{k}}{(2\pi)^3} \frac{1}{\sqrt{2E}} \sum_s \left(a^s_{\vec{k}}u^s e^{-i\mathbf{k} \cdot \mathbf{x}}   \right)\\ 
\overline{\psi}(\mathbf{x}) = \int \frac{d\vec{k}}{(2\pi)^3} \frac{1}{\sqrt{2E}} \sum_s \left(a^{s\dagger}_{\vec{k}} \overline{u}^s e^{i \mathbf{k} \cdot \mathbf{x}}\right),
\end{gathered}
\label{Fouriers}
\end{equation}
where $a^{s\dagger}_{\vec{k}}, a^s_{\vec{k}}$ are the creation and annihilation operators obeying the anticommutation relations $\left\{a^r_{\vec{k}}, a^{s\dagger}_{\vec{l}} \right\} = (2\pi)^3 \delta\left(\vec{k}-\vec{l} \,\right)\delta^{rs}$. The fermionic spinor fields $\overline{u}^s, u^s$ follow the conventional choice of basis \cite{Peskin} in spin-$z$ eigenstates, such that
\begin{equation}
u^{\uparrow}(\vec{k}) = (\sqrt{E-k_z}, 0, \sqrt{E+k_z}, 0), \quad u^{\downarrow}(\vec{k}) = (0, \sqrt{E+k_z}, 0, \sqrt{E-k_z}).
\end{equation}
To simplify the integration over the vector potential $\vec{A}$, we have assumed that its components can be approximated by their average values $\avg{A_1}, \avg{A_2}, \avg{A_3}$ over the integration volume so that they can be extracted from the integral as numbers. This volume, in which these components are nonvanishing, is fixed by the scale $d$ of the experimental apparatus used to generate the magnetic field. 

The final expression for the energy shift in our single electron state is thus:
\begin{equation}
\avg{\Delta_{\vec{A}} \mathcal{H}_{NRL}} = \frac{-|e|}{|\lambda_+|^2 + |\lambda_-|^2} \left\{\frac{m_e}{E} \left[\avg{A_1} \Re(\lambda_+^* \lambda_-) + \avg{A_2} \Im(\lambda_+^* \lambda_-) \right]+ \frac{1}{2} \avg{A_3} \left[|\lambda_+|^2 - |\lambda_-|^2 \right] \right\} + \frac{|e|\avg{A_3}}{2}\frac{|\vec{k}|}{E}.
\label{finalavgH}
\end{equation}
This expression can be compared with the corresponding one obtained in Ref. \cite{PLA2016} for the magnetic $S_z$ shift:
\begin{equation}
\avg{\Delta_{\vec{A}} S_z} = \frac{-2}{|\lambda_+|^2 + |\lambda_-|^2}\frac{|e|}{m_e} \frac{|\vec{k}|}{E} \left[\avg{A_1} \Re(\lambda_+^* \lambda_-) + \avg{A_2} \Im(\lambda_+^* \lambda_-) \right]. 
\label{S3nrl}
\end{equation}

Choosing $\vec{A} = \frac{1}{2} \vec{B} \times \vec{x}$ and assuming the spatially constant magnetic field to be oriented entirely along the $z$ axis, we find that $A_3 = 0$. Equation (\ref{finalavgH}) therefore reduces to the terms proportional to $\avg{A_1}$ and $\avg{A_2}$ only. Using characteristic values for a very weak bar magnet of $3 \times 10^{-4}$ tesla (3 gauss), an experimental apparatus of dimension $d=1$ meter, and NRL electrons with $\gamma = (1-v^2)^{-1/2} = 1.001$, we obtain $\left|\avg{\Delta_{\vec{A}} \mathcal{H}_{NRL}} \right| \lesssim 0.080$ MeV, just slightly more than $15\%$ of the electron's rest mass. We find this estimate a good validation of the perturbative expansion employed, which only retains the magnetic field to first order. Further details on numerical estimations of these effects can be found in Ref. \cite{MAGMA2017}.

In the equations above we have allowed $\lambda_+$ and $\lambda_-$ to be in general complex, but the energy and spin shifts due to the vector potential are always real. Though not required to maintain validity of the equations below, for simplicity we could limit our choice of coefficients to normalized values only and thereby constrain $|\lambda_+|^2 + |\lambda_-|^2=1$ to reflect probabilities for measuring the spin states with eigenvalues $\pm \frac{1}{2}$. With $\avg{A_3}=0$, we obtain the following relationship between the energy and spin shift expectation values:
\begin{equation}
\avg{\Delta_{B_z}\mathcal{H}_{NRL}} = \frac{m_e^2}{2|\vec{k}|} \avg{\Delta_{B_z} S_z}.
\label{HS3nrl}
\end{equation}

A major result of our calculation, given the introduction of a static magnetic field oriented along the $z$ axis, is that both energy and spin changes induced by such a field for our special single electron state with complex coefficients are proportional to the same quantity 
\begin{equation}
\Re(\lambda_+^* \lambda_-) - \Im(\lambda_+^* \lambda_-) = \Re(\lambda_+ \lambda_-^*) + \Im(\lambda_+ \lambda_-^*). 
\label{Xindex}
\end{equation}
We shall call this quantity the \textit{chirality index} in our QFT-based treatment of free electrons, for reasons that will become clear. In classical relativistic field theory, one starts from the definition of a chiral transformation \cite{Peskin} and identifies chirality as the property of how an object changes under parity (or mirror) transformations. If under these operations the object changes, it is said to be chiral. In the usual QFT approach, there are known definitions of chiral operators that are based on the $\gamma^5 \equiv i\gamma^0\gamma^1\gamma^2\gamma^3$ matrix and which act on so-called right-handed and left-handed spinors. But an explicit quantitative definition of chirality in terms of the intrinsic field components is not available. 

There exists, however, in the low-mass limit a correspondence equating the expression for chirality $\mathcal{X}$ to that of helicity:
\begin{equation} \label{chirality}
\mathcal{X} \xrightarrow{m_e \ll E} \frac{\vec{S} \cdotp \vec{k}}{\left| \,\vec{k} \,\right|},
\end{equation}
where $\vec{S}$ is the spin defined by the Pauli matrices for a particle with momentum $\vec{k}$. Our discussion of chirality will start from the assumption of a definition which reproduces the identification with helicity from Equation (\ref{chirality}). One may rightfully ask why we can use such an identification between chirality and helicity when we have derived our results in the NRL. Indeed, just as the non-relativistic Pauli equation interlocutes between the non-relativistic Schr\"{o}dinger equation and the fully relativistic Dirac equation \cite{Dirac1,Dirac2}, we shall consider the intermediate case of a free electron system in the NRL whose energy is sufficiently larger than its rest mass (but not so large as to be relativistic), according to the numerical estimates made above. Therefore the low-mass limit may be applied as a reasonable first approximation. 
Having accepted the low-mass approximation for free electrons, we obtain the following expression for the chirality in the case of an electron with momentum $\vec{k} = (0, 0, k_z)$:
\begin{equation} \label{chiS3}
\mathcal{X}_{k_z} = S_z.
\end{equation}
To verify the reasonableness of our definition in Equation (\ref{chiS3}), we have computed the expectation value of $\mathcal{X}_{k_z}$ in the state $\left |\Psi(\vec{k}) \right \rangle$ of Equation (\ref{Psi}). This is done using the Fourier expansions of the electron fields given in Equations (\ref{Fouriers}),
from which we obtain
\begin{equation}
\left \langle \mathcal{X}_{k_z} \right \rangle = \frac{\langle \Psi(\vec{k}) | \, \mathcal{X}_{k_z} \, |\Psi(\vec{k}) \rangle}{\langle \Psi(\vec{k}) | \Psi(\vec{k}) \rangle} = \frac{1}{2} \left(|\lambda_+|^2 - |\lambda_-|^2 \right).
\end{equation}
We can conclude that, in the chosen definition, $\langle \mathcal{X}_{k_z} \rangle$ satisfies the intuitive understanding of chirality as a difference between the moduli of right-handed and left-handed spin polarization coefficients of the electron state. 

We have now in a QFT framework an operative calculational tool, in terms of the fields $\psi$ and $\overline{\psi}$, for the chirality $\mathcal{X}_{k_z}$ and for the energy $\mathcal{H}$ of the electron state. 
For a magnetic field along the $z$ axis, the result for the average magnetic effect on the chirality of electron state (\ref{Psi}), in our approximations, is identical to the computed $S_z$ shift of Equation (\ref{S3nrl}): $\left \langle\Delta_{\vec{A}} \mathcal{X}_{k_z} \right \rangle = \left \langle\Delta_{\vec{A}} S_z \right \rangle$.
Most notably, we can say that under these conditions the magnetic effects on the chirality and on the energy of free electrons are proportional to the same intrinsic electron property that we defined as the \textit{chirality index}. An immediate consequence of Equations (\ref{HS3nrl}) and (\ref{chiS3}) is that there exists a special quantity of our considered system whose expectation value remains constant upon introduction of the magnetic field:
\begin{equation} \label{CEC}
\left\langle \Delta_{B_z} \left(\mathcal{H}_{NRL} - \frac{m_e^2}{2 k_z} \mathcal{X}_{k_z} \right) \right\rangle = 0.
\end{equation}
We say that there appears to be \textit{chirality-energy conversion} in the system under the effect of a static magnetic field that is similar to the spin-orbital angular momentum conversion predicted in QFT \cite{PLA2016} and confirmed experimentally \cite{SpinOAM1,SpinOAM2,SpinOAM3}. 

Another possible way of interpreting Equation (\ref{CEC}) is to see it as an indicator of a special magnetic symmetry of the free electron system. To extend our analysis, we consider the case of $N$ free electrons, each with the same energy $E$. It seems to us reasonable that a magnetic field would act independently on each of the $N$ electrons. Therefore, for each $j$th electron, one would find the following expression, where the NRL is implied:
\begin{equation} \label{jth}
\left \langle\Delta_{B_z} \mathcal{X}_{k_z} \right \rangle_j = \frac{2k_z}{m_e^2} \left \langle \Delta_{B_z} \mathcal{H} \right \rangle_j.
\end{equation}
It is then possible to define mean values for the system of free electrons, 
\begin{equation}
\begin{gathered}
\left \langle\Delta_{B_z} \mathcal{X}_{k_z} \right \rangle_{sys} = \frac{1}{N} \sum_j \left \langle\Delta_{B_z} \mathcal{X}_{k_z} \right \rangle_j \\
\left \langle \Delta_{B_z} \mathcal{H} \right \rangle_{sys} = \frac{1}{N} \sum_j \left \langle\Delta_{B_z} \mathcal{H} \right \rangle_j,
\end{gathered}
\end{equation}
such that by using Equation (\ref{jth}) we obtain the following analogous relation for the entire collection:
\begin{equation} \label{sys}
\left \langle\Delta_{B_z} \mathcal{X}_{k_z} \right \rangle_{sys} = \frac{2k_z}{m_e^2} \left \langle \Delta_{B_z} \mathcal{H} \right \rangle_{sys}.
\end{equation}
If it is possible to measure the energy shift of free electrons induced by a magnetic field, one immediately derives from Equation (\ref{sys}) a QFT prediction for the change in chirality of the system, which could also be experimentally measured. Thus we would be able to test the validity of our theory. 

A key question that remains to be considered is the possible relevance of our results to macroscopic states in quantum optics, condensed matter, and biological physics, where collections of free or quasi-free electrons may be described in the formalism above. Indeed, though the description of such macroscopic states is complex, recent experimental studies \cite{Naaman} indicate that electrons transmitted through chiral molecules may be filtered according to their spin state. Furthermore, it has long been known that a sensitive dependence exists between the chirality of  crystals and low-energy fluctuations introduced by perturbing the crystallization solution \cite{Sci1990}. 

Such sensitive relationships between biological function and chirality of the underlying spin state are manifest with both free and bound electron states. Several articles since 2005 have reported effects of weak magnetic fields on the rate of enzymatic synthesis of adenosine triphosphate \cite{HorePNAS} and reactive oxygen species \cite{MartinoQB} by the flipping of electron spins in a quantum-coherent fashion. We have shown theoretically \cite{JTB} that palindromic DNA complexes of defined chirality conserve parity and are essential to the symmetric recruitment of energy by certain enzymes for the formation of DNA double-strand breaks. These evidences all point to the existence of an elaborate hierarchy of order connecting the spin states of electron systems to manifestations of \textit{chirality-energy conversion} at multiple physical scales. Our group of dedicated colleagues has already started pursuing experimental verification of this hypothesis.\\

\textbf{Data accessibility.} This work does not have any experimental data.\\

\textbf{Competing interests.} We have no competing interests. \\

\textbf{Author's contributions.} PK completed all derivations and calculations, with numerical estimates, and wrote the paper. The author gives final approval for publication. \\

 \textbf{Acknowledgements.} This manuscript was motivated by discussions with C. Verzegnassi of the University of Udine and of the Association for Medicine and Complexity (Trieste, Italy). His insights were essential to its development.\\
 
 \textbf{Funding.} PK would like to acknowledge support from the U.S. -- Italy Fulbright Commission and the Whole Genome Science Foundation. \\


\begin{thebibliography}{99}
\bibitem{PLA2016}
P. Kurian and C. Verzegnassi. Phys. Lett. A \textbf{380,} 394-396 (2016).
\bibitem{Ultramicro}
P. Schattschneider and J. Verbeeck. Ultramicroscopy \textbf{111,} 1461-1468 (2011).
\bibitem{Peskin}
M. E. Peskin and D. V. Schroeder. \textit{An Introduction to Quantum Field Theory}. Perseus, 1995.
\bibitem{MAGMA2017}
C. Verzegnassi, R. Germano, and P. Kurian. J. Magn. Magn. Mater. \textbf{449,} 482-484 (2018).
\bibitem{Dirac1}
P. A. M. Dirac. Proc. R. Soc. Lond. A \textbf{117,} 610-624 (1928).
\bibitem{Dirac2}
L. L. Foldy and S. A. Wouthuysen. Phys. Rev. \textbf{78,} 29-36 (1950).
\bibitem{SpinOAM1}
V. Grillo et al. Phys. Rev. Lett. \textbf{114,} 034801 (2015).
\bibitem{SpinOAM2}
E. Karimi et al. Phys. Rev. Lett. \textbf{108,} 044801 (2012).
\bibitem{SpinOAM3}
G.-Y. Guo, S. Maekawa, and N. Nagaosa. Phys. Rev. Lett. \textbf{102,} 036401 (2009).
\bibitem{Naaman}
V. Kiran, S. R. Cohen, and R. Naaman. J. Chem. Phys. \textbf{146,} 092302 (2017).
\bibitem{Sci1990}
D. K. Kondepudi, R. J. Kaufman, and N. Singh. Science \textbf{250,} 975-976 (1990).
\bibitem{HorePNAS}
P. J. Hore. Proc. Natl. Acad. Sci. USA \textbf{109,} 1357-1358 (2012).
\bibitem{MartinoQB}
R. J. Usselman, C. Chavarriaga, P. R. Castello, M. Procopio, T. Ritz et al. Sci. Rep. \textbf{6,} 38543 (2016).
\bibitem{JTB}
P. Kurian, G. Dunston, and J. Lindesay. J. Theor. Bio. \textbf{391,} 102-112 (2016).
\end{thebibliography}
\end{document}